# Silicon Diphosphide (SiP$_2$) and Silicon Diarsenide (SiAs$_2$): Novel Stable 2D Semiconductors with High Carrier Mobilities and Promising for Water Splitting Photocatalysts

Fazel Shojaei *[a], Bohayra Mortazavi [b], Xiaoying Zhuang [b] and Maryam Azizi *[a]

[a] School of Nano Science, Institute for Research in Fundamental Sciences (IPM), P. O. Box: 19395-5531, Tehran, Iran

[b] Institute of Continuum Mechanics, Leibniz Universitöt Hannover, Appelstraße 11, 30157 Hannover, Germany

**Abstract**

Two dimensional (2D) semiconducting light absorbers, have recently considered as promising components to improve the efficiency in the photocatalytic hydrogen production via water splitting. In this work, by employing density functional theory computations, we introduced novel SiX$_2$ (X = P, As) nanosheets in tetragonal (penta-) and orthorhombic (rec-) phases, as promising light absorber semiconductors for overall water splitting. The predicted nanomembranes exhibit good mechanical, dynamical and thermal stabilities. They also show small cleavage energies in the range of 0.31 J/m$^2$ to 0.39 J/m$^2$, comparable to that of the graphene and thus suggesting the feasibility of their experimental exfoliation. Notably, predicted monolayers are semiconductors with indirect band gaps of 2.65 eV for penta-SiP$_2$, 2.35 eV for penta-SiAs$_2$, 1.89 eV for rec-SiAs$_2$, and a direct band gap of 2.21 eV for rec-SiP$_2$. These nanomaterials however show relatively large interlayer quantum confinement effects, resulting in smaller band gap values for bilayer lattices. We observed a huge difference between the electron and hole mobilities for penta-SiP$_2$ and rec-SiAs$_2$ monolayers and highly directional dependent electron and hole mobilities in rec-SiP$_2$, yielding an effective separation of photogenerated charge carriers. Remarkably, these novel nanomembranes show strong absorption in the visible region of light as well as suitable band edge positions for photocatalytic water splitting reaction, specifically under neutral conditions.

# 1. Introduction

Production of hydrogen via photocatalytic water splitting using the sunlight is considered as a promising solution to our rising demand for clean and renewable energy resources.[1, 2, 3] Generally, a photocatalytic process involves three distinct steps of light absorption, separation and transport of photo-generated charge carries in the bulk, and electrochemical catalysis on the surface of the light absorber.[4] It can be seen that the core of a photocatalytic hydrogen production process is a light-absorber semiconductor that converts the sunlight into the chemical energy. The electronic structure of a semiconductor plays an important role in determining its possible photocatalytic activity for water splitting. A suitable semiconductor for splitting water is expected to have a sufficient band gap energy between 1.23 eV to 3 eV. In addition, its band edge positions must also straddle the water redox potentials. Several other criteria a semiconductor must fulfill to become an efficient and useful photocatalyst for practical applications.[5] Therefore, only a few of the known semiconducting materials are suitable for this purpose. Since the first report on splitting water over $TiO_2$ electrodes in 1972, [2] a significant amount of research efforts has been devoted to explore other novel photocatalysts. Some other semiconductors such as ZnO, CdSe, CdS, $WO_3$ , and $SrTiO_3$ were also identified to exhibit photocatalytic activity for water splitting.[6, 7] The efficiency of these bulk photocatalysts is limited by the absence of active reaction sites and low conversion efficiency due to the charge carrier recombination.[5] Therefore, finding stable and efficient photocatalyst materials still remains very challenging.

Two dimensional (2D) materials which have been a hot topic in last decade, present a broad range of tunable electronic and optical properties that make them promising candidates for nanoelectronics,[8, 9, 10, 11] energy conversion/storage,[12, 20, 13, 14] and catalysis.[15, 16, 17] Several experimental studies have shown that 2D material-based photocatalysts exhibit highly improved photocatalytic activities compared to the traditional bulk materials.[18, 19, 20, 21] This is because of the unique advantages of 2D material photocatalysts over their bulk counterparts: exceptionally large surface area that results in more surface active sites, and better separation of photo-generated electrons and holes due to the atomic thickness of the sheet so that the charge carriers can reach the surface active sites before they recombine.[5] The photocatalytic activities of some 2D materials such as metal oxides (ZnO,[19] $TiO_2$ ,[22]), metal dichalcogenides ($SnS_2$ ,[18] $MoS_2$ ,[23] $TiS_2$ [24]), and metal-free semiconductors (phosphorene,[25] g-$C_3N_4$ ,[17] antimonene,[26]) have been experimentally investigated in recent years. In addition, several other 2D materials have been theoretically predicted to exhibit photocatalytic activity for water splitting.[27, 28, 29, 30, 31, 32, 33, 34, 35, 36, 37, 38, 39, 40]

Binary IV-V compounds with the chemical formulas of IVV (GeP, GeAs, SiP, SiAs),[41, 42] $IVV_2$ ,($SiP_2$ , $SiAs_2$ , $GeP_2$ , $GeAs_2$ )[43, 44, 45, 32] $IVV_3$ ($GeP_3$ ,$SnP_3$ ), [42, 46] and $IVV_5$ ($GeP_5$ ),[47] are emerged as another interesting class of semiconducting 2D materials. Our literature review have shown that IVV, $IVV_2$, and $IVV_3$ compounds exhibit polymorphism. As an example, theoretical calculations suggested the possible experimental synthesis of a hexagonal phase with P $\bar{6}m2$ space group for IVV compounds, which is marginally less stable than the experimentally observed 2D low symmetry monoclinic phase.[48, 49] As another example, a layered tetragonal phase with famous penta-graphene-like structure has been suggested for $IVV_2$ [34, 45, 32] compounds in addition to experimentally identified non-Van der Waals pyrite phase and 2D layered low symmetry orthorhombic structures.[43, 44] Results of joint experimental and theoretical studies confirmed the photocatalytic activity of mono- to few-layer GeP [50] and GeAs [51] materials for water splitting. It has been also predicted that penta-$GeP_2$ and orthorhombic $SiAs_2$ are promising candidates for photocatalytic water splitting and $CO_2$ reduction.[45, 52]

In this work, inspired by some of group IV-V 2D materials which have been theoretically predicted and experimentally confirmed to exhibit photocatalytic activity for water splitting reaction, we investigated the electronic structures, charge carrier mobilities, optical properties, and band edge positions of polymorphic $SiX_2$ (X = P, As) nanosheets. Two layered phases with tetragonal (penta-) and orthorhombic (rec-) crystal structures are identified for the nanosheets. The monolayers are found to be semiconductors with indirect band gaps of 2.65 eV for penta-$SiP_2$, 2.35 eV for penta-$SiAs_2$, 1.89 eV for rec-$SiAs_2$, and a direct band gap of 2.21 eV for rec-$SiP_2$ systems. They show strong absorption in the visible region as well as suitable band edge positions for photocatalytic water splitting reaction, specifically under neutral condition. We observed a huge difference between the mobilities of electrons and holes in penta-$SiP_2$ and rec-$SiAs_2$ monolayers and highly directional electron and hole mobilities in rec-$SiP_2$ monolayer, resulting in an effective separation of photogenerated electrons and holes. In short, our calculations reveal that $SiX_2$ (X = P, As) nanosheets would be promising photocatalysts for water splitting under the visible sunlight.

## 2. Computational methods

All the structural relaxations and electronic structure calculations were performed using the periodic density functional theory (DFT) as implemented in VASP.[53, 54] The electron-ion interactions were described using the projector augmented wave (PAW) method, which is primarily a frozen-core all-electron calculation.[54] The weak van der Waals interactions in monolayer, few-layer, and bulk $SiX_2$

(X = P, As) structures were taken into account employing Grimme's correction of D3 (PBE-D3).[55] For structure optimization, atoms are relaxed in the direction of the Hellmann-Feynman force using the conjugate gradient method with an energy cut-off of 500 eV until a stringent convergence criterion (= 0.001 eV/Å) is satisfied. HSE06 hybrid functional is employed for high-accuracy electronic structure calculations.[56] Proper Γ-centered κ-point meshes are used to sample the Brillouin zones of our 2D and bulk systems. We define the 2D layer to lie on the X-Y plane so that the Z axis is parallel to the c axis. For the 2D systems, a sufficiently large vacuum space of 15 Å is maintained along the Z direction to ensure that no appreciable interaction occurs between two adjacent supercells.

To examine the dynamic stability of the monolayers, we calculated their phonon band structures by employing density functional perturbation theory (DFPT) method implemented in the PHONOPY package [57, 58] using 4×4×1 and 2×6×1 super-cells for the penta and rectangular lattices, respectively. To further verify the thermal stability of a specific phase, Nose-Hoover molecular dynamics (MD) simulations are performed in constant NVT systems with a time step of 1.0 fs and Γ-point sampling.[59, 60, 61] Mechanical properties were evaluated by conducting the uniaxial tensile simulations. As it is clear, penta structures are completely isotropic along the two perpendicular directions. Because of the anisotropic nature of the rectangular lattices, the tensile loading conditions were applied along the x and y directions, as depicted in Fig. 1. For the uniaxial tensile simulations, the periodic simulation box size along the loading direction was increased gradually with a fixed strain step, using a 7×7×1 Monkhorst-Pack [1] κ-point mesh size. After applying the loading strain and accordingly rescaling the atomic positions, the simulation box size along the sheet perpendicular direction of the loading was adjusted to reach negligible stress.

In a 2D system, the carrier mobility can be calculated from the deformation energy approximation:[62]

$$\mu_{2D} = \frac{eh^3 C_{2D}}{KT m_e^* m_d (E_1^i)^2}$$

where $C_{2D}$ is the elastic modulus of the longitudinal strain in the transport direction, m* is the effective mass of the carrier in the same direction, $m_d$ is the average effective mass in the two directions given by $m_d = \sqrt{m_x^* m_y^*}$ and $E_1^i$ mimics the deformation energy constant of the carrier due to phonons for the i-th edge band along the transport direction through the relation:

$$E_1^i = \frac{\Delta E_i}{\Delta L/L_0}$$

, where ΔE i and ΔL/L₀ represent the energy change of the i-th band and lattice dilation, respectively.

## 3. Results and discussion

We first investigate the polymorphism in bulk $SiX_2$ (X = P, As) compounds. Worthy to note that two distinct phases are experimentally identified for bulk $SiX_2$, namely, a pyrite-type and an orthorhombic phase. The pyrite phase has a non-Van der Waals crystal structure with Pa3 (No. 205) space group, while the orthorhombic phase shows a 2D layered structure with Pbam (No. 55) space group. [43, 44] Recently, a new 2D layered phase with a pentagonal bonding pattern has been also predicted for group IV-V materials with $IVV_2$ chemical formula. This phase has a tetragonal crystal structure with $P-42_1m$ (No. 113) space group and it is composed exclusively of fused $Si_2X_3$ pentagons.[34, 45, 32] Fig. 1 shows the chemical structures of these three aforementioned phases. Both layered orthorhombic and tetragonal phases obey the octet rule, in which each Si atom is four-coordinated by four X atoms and each X atom is three coordinated by two Si atoms and one X atom. However, the octet rule in non-Van der Waals pyrite phase is violated so that each Si atom is six-coordinated by six X atoms and each X atom is four coordinated by three Si atoms and one X atom.

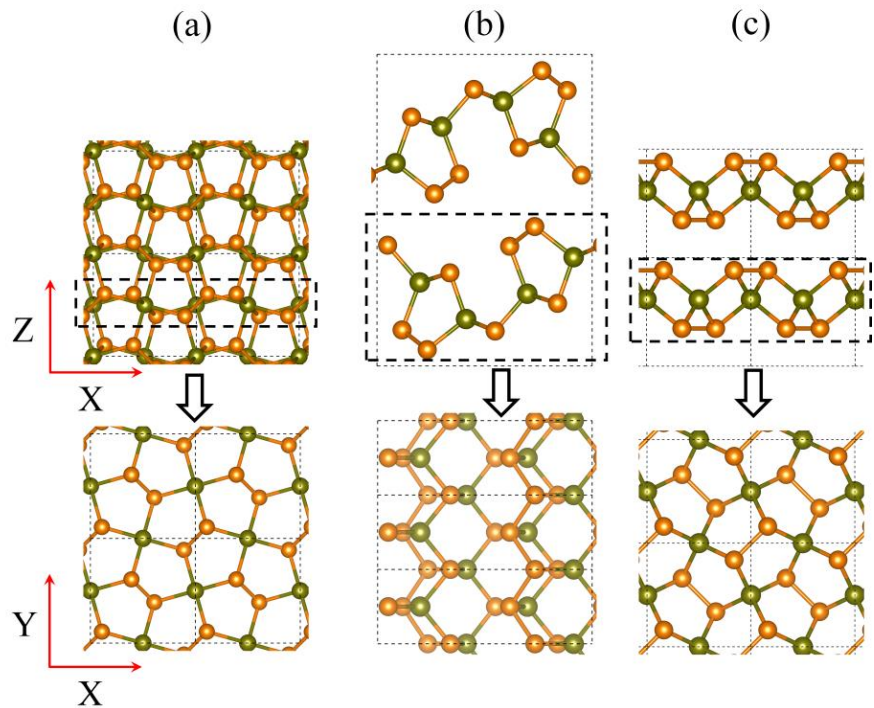

**Figure 1** Different views for the chemical structures of bulk $SiX_2$ in (a) pyrite, (b) orthorhombic, and (c) tetragonal phases. The top view of an isolated monolayer of each phase, enclosed in the dashed box, is also shown. Orange and green colors represent silicon and X (P,As) atoms, respectively.

In Table. 1 we summarized the structural and energetic properties of the three phases. According to the Table, for SiP$_2$, the pyrite structure is the energetically most stable phase, which is lower than the orthorhombic and tetragonal phases by 36 and 57 meV/atom, respectively. In the case of SiAs$_2$, the orthorhombic phase is found to be the lowest energy structure, and the pyrite phase is 83 meV/atom higher in energy. Moreover, the tetragonal phase energetically lies between the orthorhombic and pyrite phases. In fact, the constituent layers of a tetragonal SiX$_2$ can be fabricated by exfoliation of the corresponding pyrite structure. For better understanding, the top view of an isolated layer of each phase (enclosed in dashed boxes) is also shown in Fig. 1. It can be seen that the monolayers obtained from pyrite and tetragonal phases have similar bonding patterns. However, the monolayer from tetragonal phase is significantly more buckled than the one exfoliated from bulk pyrite structure. Our solid-state nudged elastic band (G-SSNEB) calculations [65] show that the exfoliated layer can be instantly transform to the layer from the tetragonal phase with no energy barrier. Therefore, one might naturally conclude that the tetragonal phase could be also experimentally synthesized under proper conditions. This idea is supported by the fact the exfoliation of non-Van der Waals materials has been recently realized using liquid phase exfoliation method for iron ore hematite (α-Fe$_2$O$_3$) and titanate ore ilmenite (FeTiO$_3$). [66, 67] Inspired by the several merits of 2D materials in various fields of science and technologies, we will limit our further discussions to the layered tetragonal and orthorhombic phases. In two dimension, tetragonal and orthorhombic phases have square and rectangular lattice structures, respectively. Henceforth, otherwise explicitly stated, we call systems with square and rectangular lattice structures penta and rec, respectively.

**Table 1** PBE-D3 calculated structural and energetic properties of the three identified phases for bulk SiX$_2$. For comparison the experimentally obtained lattice constants of pyrite and orthorhombic phases of each compound are also shown.

| SiX$_2$ | | | (a, b, c)[1] | E$_{rel}$ (eV/atom)[2] | d$_{Si-X}$[3] | d$_{X-X}$[4] | Layer Thickness[5] | Inter-layer Distance[6] |
|---|---|---|---|---|---|---|---|---|
| X = P | pyrite | Cal | (5.71,5.71,5.71) | 0 | 2.40 | 2.16 | - | - |
| | | Exp[63] | (5.70, 5.70, 5.70) | - | - | - | - | - |
| | orthorombic | Cal | (9.91,14.40,3.45) | 0.036 | 2.27 | 2.27 | 5.69 | 1.51 |
| | | Exp[43] | (10.08,13.98,3.436) | - | - | - | - | - |
| | | 1L* | (10.03, 3.44) | 0 | 2.26 | 2.26 | 5.58 | - |
| | pentagonal | Cal | (4.86,4.86,5.07) | 0.0567 | 2.27 | 2.22 | 2.72 | 2.35 |
| | | 1L* | (4.80, 4.80) | 0.063 | 2.27 | 2.23 | 2.78 | - |
| X = As | pyrite | Cal | (6.07,6.07,6.07) | 0.084 | 2.54 | 2.44 | - | - |
| | | Exp[63] | (6.02, 6.02, 6.02) | - | - | - | - | - |
| | orthorombic | Cal | (10.14, 14.90, 3.69) | 0 | 2.41-2.45 | 2.51 | 6.17 | 2.22 |
| | | Exp[64] | (10.37,14.53,3.64) | - | - | - | - | - |
| | | 1L* | (10.026, 3.45) | 0 | 2.40-2.46 | 2.50 | 5.58 | - |
| | pentagonal | Cal | (5.10,5.10,5.43) | 0.034 | 2.40 | 2.49 | 2.98 | 2.46 |
| | | 1L* | (4.78, 4.78) | 0.064 | 2.40 | 2.50 | 2.78 | - |

[1]PBE-D3 optimized lattice constants. [2] Relative energy with respect to the most stable phase of each compound. [3] Si-X bond length bond length in Å.[5]The layer thickness is defined by the difference between the maximum and minimum Z-coordinates of atoms to the same layer. [6] The inter-layer distance is defined by the difference between the maximum and minimum Z-coordinates of the upper layers.

Before we start studying the electronic and optical structures of these materials we need to make sure whether they are stable. As a common approach to examine the stability of novel 2D materials, we then evaluate the dynamical stability of the predicted nanosheets by calculating the phonon dispersion relations along the high symmetry directions of the first Brillouin zone. The acquired phonon dispersion for the SiP$_2$ and SiAs$_2$ monolayers with rectangular (rec-) and square (penta-) lattices are shown in Fig. 2. Notably, for the predicted monolayers with the penta structure, the dynamical matrix is completely free of any imaginary eigenvalues, which would appear as negative branches, and thus confirming the dynamical and structural stability of these novel materials. For these nanosheets, two of the three acoustic modes present linear dispersion, and the remaining one exhibits a quadratic curvature, which are the characteristic features of the 2D materials. On the other side, the rectangular counterparts also show two linear acoustic modes around the gamma point. However, in these cases the acoustic mode with the quadratic dispersion shows a negligible pit shaped negative part in the Γ-X path. The corresponding modes along the Γ-S and Γ-Y are however completely free of imaginary frequencies. The slight U shaped feature in the beginning of the Γ-X path is a signature of the flexural acoustic mode, which is usually hard to converge in 2D materials [68] and might be removed by increasing the accuracy of calculations, upon the increasing the super-cell size or k-point grid, which are computationally highly demanding. On the other hand, in the cases of multilayer structures or supporting over a substrate, these slight U shaped negative branches normally vanishes. According to the phonon dispersion results, it can be concluded that the predicted nanosheets are dynamically stable. The thermal stabilities of the monolayers are also examined by performing ab initio MD simulations (AIMD) at 300 K employing 3×3 and 2×4 supercells for penta- and rec-SiX$_2$ monolayers, respectively. Fig. S1 shows their chemical structures after 5 ps. It can be clearly seen that their geometrical integrity retain and chemical bonds are only slightly elongated, confirming their thermal stabilities.

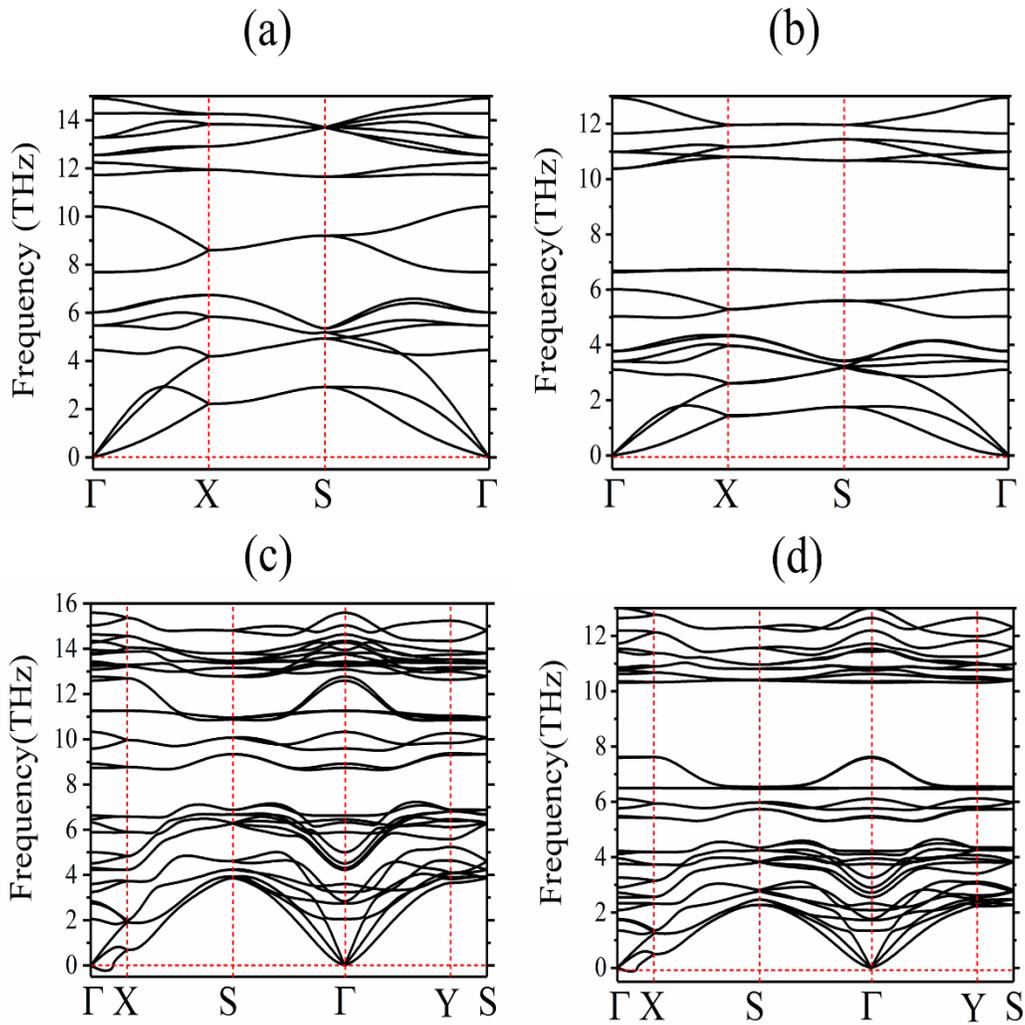

**Figure 2** Phonon dispersion spectra of (a) penta-SiP$_2$, (b) penta-SiAs$_2$, (c) rec-SiP$_2$ and (d) rec-SiAs$_2$ monolayers.

We next study the mechanical properties of the predicted nanosheets on the basis of uniaxial tensile simulations. The computed uniaxial stress-strain responses of penta- and rec-SiX$_2$ monolayers elongated along X and Y directions are illustrated in Fig. 3. Likely to other conventional materials, the stress-strain curves exhibit initial linear responses, corresponding to the linear elasticity. Results shown in Fig. 3 reveal highly anisotropic elastic response for rectangular lattices, in contrast to the isotropic nature of square counterparts. According to our results, the elastic modulus of penta-SiP$_2$ and penta-SiAs$_2$ were found to be 67.9 and 47.5 N/m, respectively. The elastic modulus of rec-SiP$_2$ along the Y than X directions were predicted to be 128.0 and 24.3 N/m, respectively, and for rec-SiAs$_2$ the corresponding values were estimated to be 115.2 and 23.2 N/m, respectively. As it is clear, SiP$_2$ allotropes show higher rigidity as compared with SiAs$_2$ counterparts. A similar anisotropic elastic modulus were observed for α-phosphorene (88.02 N/m for zigzag direction and 26.16 N/m for armchair direction ).[69] On the other hand, maximum tensile strength of rec-SiP$_2$ along the Y than X directions were estimated to be 14.5 and 8.2 N/m, respectively, and those for rec-SiAs$_2$ monolayer were calculated to be 11.8 and 6.6 N/m, respectively. Notably, while the anisotropicity in the elastic modulus of rectangular lattices are around the ratio of 5, for the tensile strength this ration turns to be less than 2. The tensile strength of SiP$_2$ and SiAs$_2$ nanosheets with penta structures were found to be very close, 9.6 and 9.1 N/m, respectively. As an interesting observation, although the SiAs$_2$ lattices are softer and less strong than the SiP$_2$ counterparts, and they can show load bearing at higher strain levels and thus they are more stretchable. Notably, along the X direction, the rectangular lattices can yield super stretchability, as the tensile strengths occur at strain levels around 0.45, which is almost twice as that of the pristine graphene, about 0.27.[70]

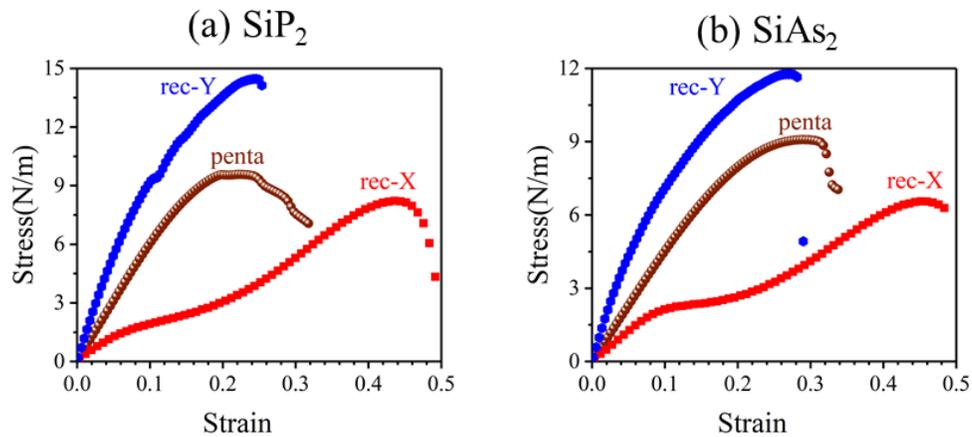

**Figure 3** Uniaxial stress-strain responses of penta- and rec-SiX$_2$ monolayers. The structures with rectangular lattices are studied in both X and Y directions.

The feasibility of isolation of a monolayer or few-layer nanosheets from their layered bulk crystals is of great fundamental and practical importance. We quantified the exfoliation ease level of SiX$_2$ (X = P, As) monolayers by estimating their cleavage energies from a 6 layer slab of each phase, approximated to be as a model of the bulk structures. As shown in Fig. 4 the calculated cleavage energies are 0.39 J/m$^2$ for penta-SiP$_2$ and 0.35 J/m$^2$ for penta-SiAs$_2$ monolayers, which are slightly larger than those of 0.31 J/m$^2$ for rec-SiP$_2$ and 0.35 J/m$^2$ for rec-SiAs$_2$ monolayers. Such small values are close to the experimentally calculated cleavage energy of graphene (0.37 J/m$^2$) [71] and much lower than DFT estimated values for GeP$_3$ (1.14 J/m$^2$), [72] NaSnP (0.81 J/m$^2$), [73] and the experimentally realized Ca$_2$N monolayer (1.09 J/m$^2$), [74] indicating the feasibility of fabricating these monolayers from their bulks in laboratory. It is worth noting that a few explicit approximation were used in calculating the exfoliation energy by using the slab method. In 2018, Jung et al. [75] proposed another method for calculating the exfoliation energy of 2D materials, which was shown to be exact and much less computationally expensive compared to the slab method. We also calculated the exfoliation energies of SiX$_2$ monolayers using the newly proposed method and found that the obtained values (penta-SiP$_2$ (0.39 J/m$^2$), penta-SiAs$_2$ (0.35 J/m$^2$), rec-SiP$_2$ (0.30 J/m$^2$) and rec-SiAs$_2$ (0.33 J/m$^2$)) are in very good agreement with those calculated using the slab method.

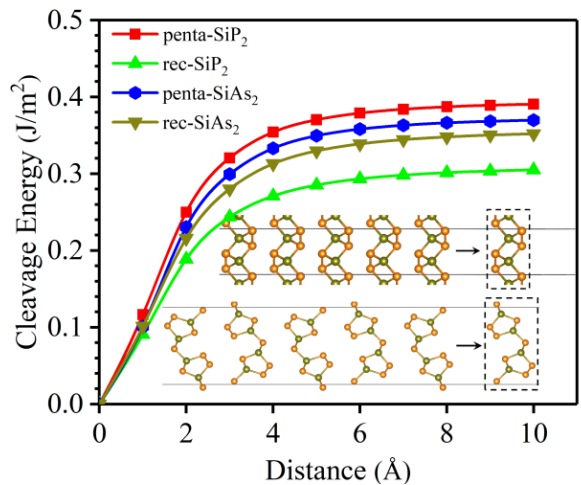

**Figure 4** Cleavage energy as a function of separation distance between the SiX$_2$ monolayer and a five-layer slab.

Having a sufficiently large band gap (1.23 eV< E$_g$ <3 eV) is the primary criterion for a good photocatalyst. To examine the validity of this criterion for SiX$_2$ (X = P, As) monolayers, we next investigated their electronic properties using HSE06 functional because as it is well-known, standard DFT underestimates the band gaps of semiconductors. Fig. 5 presents the electronic band structures, partial density of states (PDOS), and charge density distributions at the valance band maximum (VBM) and conduction band minimum (CBM) for $SiX_2$ monolayers. We first investigate monolayers with tetragonal structure. It can clearly be seen that both monolayers are semiconducting with indirect band gaps of 2.65 eV for penta-$SiP_2$ and 2.35 eV for penta-$SiAs_2$ (Fig. 5 (a) and (b)). The band gap values are appreciably larger than that of α-phosphorene (1.53 eV),[76] while they are close to that calculated for β-arsenene (2.49 eV).[77] The VBM for penta-$SiAs_2$ is located at X point whereas for penta-$SiP_2$, it is slightly off-X (0.02 eV higher than X point ) at X' = 0.464 $κ_x$/2+ 0 $κ_y$/2 on Γ-X path, where $κ_x$ and $κ_y$ are the magnitudes of the reciprocal lattice vectors along the X and Y directions. For both materials, the CBM occurs at S' = 0.214 $κ_x$/2 + 0.214 $κ_y$/2 on S-Γ path. Our analysis of PDOS and charge density distributions indicates that for penta-$SiP_2$, the VBM mainly consists of P 3p orbitals, representing anti-bonding $(P-P)_{π*}$ states. On the other hand, VBM of penta-$SiAs_2$ is made of As 4p and Si 3p orbitals, showing bonding like $(As-As)_π$ and $(As-Si)_σ$ states in addition to anti-bonding $(As-As)_{π*}$. It is found that for both materials, CBM is also almost equally contributed by P/As 3p/4p and Si 3p orbitals, forming antibonding $(P-P)_{σ*}$ and $(As-As)_{σ*}$ states.

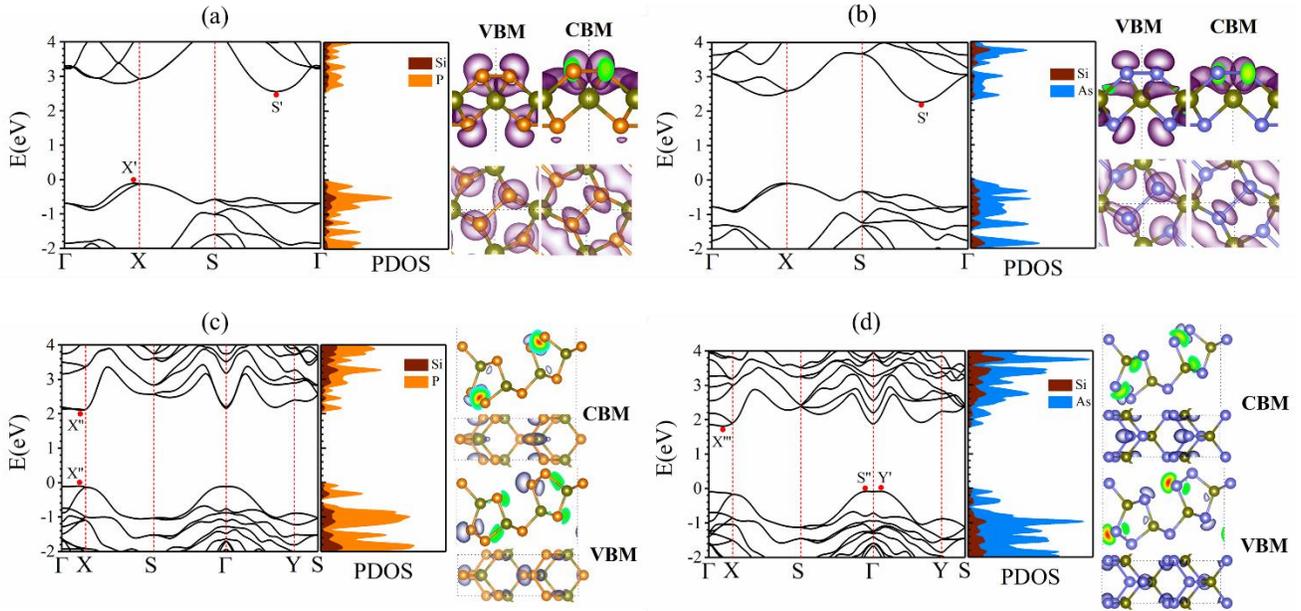

**Figure 5** HSE06 band structures and the corresponding partial density of states (PDOS) of (a) penta-$SiP_2$, (b) penta-$SiAs_2$, (c) rec-$SiP_2$, and (d) rec-$SiAs_2$ monolayers. Charge density distributions of VBM and CBM are also shown for each system. Solid red circles indicate the positions of k-points that VBM or CBM occurs.

The structure and bonding patterns in different allotropes of a material greatly influence the electronic structure and in turn band gap and optical properties. According to our HSE06 band structures, rec-$SiX_2$ (X = P, As) monolayers are also semiconductors, but have a direct band gap of 2.21 eV for rec-

SiP$_2$ and an indirect band gap of 1.89 eV for rec-SiAs$_2$ . The calculated band gaps are about 0.45 eV smaller than those of their corresponding penta monolayers. Due to the anisotropy in structure, the monolayers exhibit strongly anisotropic band dispersion along the Γ-X and Γ-Y directions near the edge of valance and conduction bands. Both VBM and CBM of rec-SiP$_2$ are located at X'' = 0.357 κ$_x$/2 + 0 κ$_y$/2 on Γ-X path. Interestingly, another direct gap of rec-SiP$_2$ monolayer at Γ point (2.25 eV) is only 0.04 eV larger than its true band gap. The calculated band gap and transition k-points are in very good agreement with a previous data (2.25 eV) obtained using HSE-Wannier function method.[52] The HSE06 band structures for rec-SiAs$_2$ in Fig. 5(d) indicates that the VBM is almost doubly degenerates at k-points of Y' = 0 κ$_x$/2+ 0.071 κ$_y$/2 and S'' = 0.036 κ$_x$/2 + 0.036  κ$_y$/2 along Γ-Y and Γ-S paths, respectively, while the CBM occurs at X'''= 0.286 κ$_x$/2 + 0  κ$_y$/2 on Γ-X path. The direct gap at Γ point is only 0.06 eV larger than its actual indirect gap. The calculated band gap and transition k-points are in good agreement with a previous data (2.07 eV) obtained using HSE06 functional. Our careful analysis of PDOS and charge density distributions indicates that for both compounds, the VBM and CBM are composed of localized P/As 3/4p states with no appreciable inter- and intra-cell interactions along the X direction, leading to almost dispersionless edge states along the Γ-X path.

It is well-known that band gap energy and sometimes even the nature of band gap alter when the monolayers are stacked on top of each other to form bilayer or multilayers. We calculated the electronic band structures of SiX$_2$ (X = P, As) bilayers to evaluate the influence of decreasing quantum confinement on the electronic properties of the monolayers. The same stacking pattern as in bulk SiX$_2$ were used for constructing the bilayers. Fig. 6 shows the HSE06 band structures and charge density distributions at VBM and CBM for SiX$_2$ (X= P, As) bilayers. It is found that SiX$_2$ (X = P, As) bilayers are also semiconductors with indirect band gaps of 2.30 eV for penta-SiP$_2$ bilayer, 1.98 eV for penta-SiAs$_2$ bilayer, and 1.51 eV for rec-SiAs$_2$ , and a quasi-direct gap of 2.02 eV for rec-SiAs$_2$ bilayer. The calculated band gaps are 0.35 eV, 0.37 eV, 0.38 eV, and 0.19 eV smaller than those of the corresponding monolayers, indicating strong quantum confinements in penta-SiP$_2$, penta-SiAs$_2$ bilayer, and rec-SiP$_2$ compounds. The transition k-points for indirect gap bilayers are the same as those of the corresponding monolayers, except for the VBM of penta-SiP$_2$ bilayer which is lies on Γ-X path. Different than the rec-SiP$_2$ monolayer, the VBM of rec-SiP$_2$ bilayer is slightly off-X, lies on the Γ-X path, while the CBM is located at X point. However, the gap at κ$_{VBM}$ is only 1 meV smaller than the indirect gap, indicating that rec-SiP$_2$ bilayer can be considered as a quasi-direct gap material.

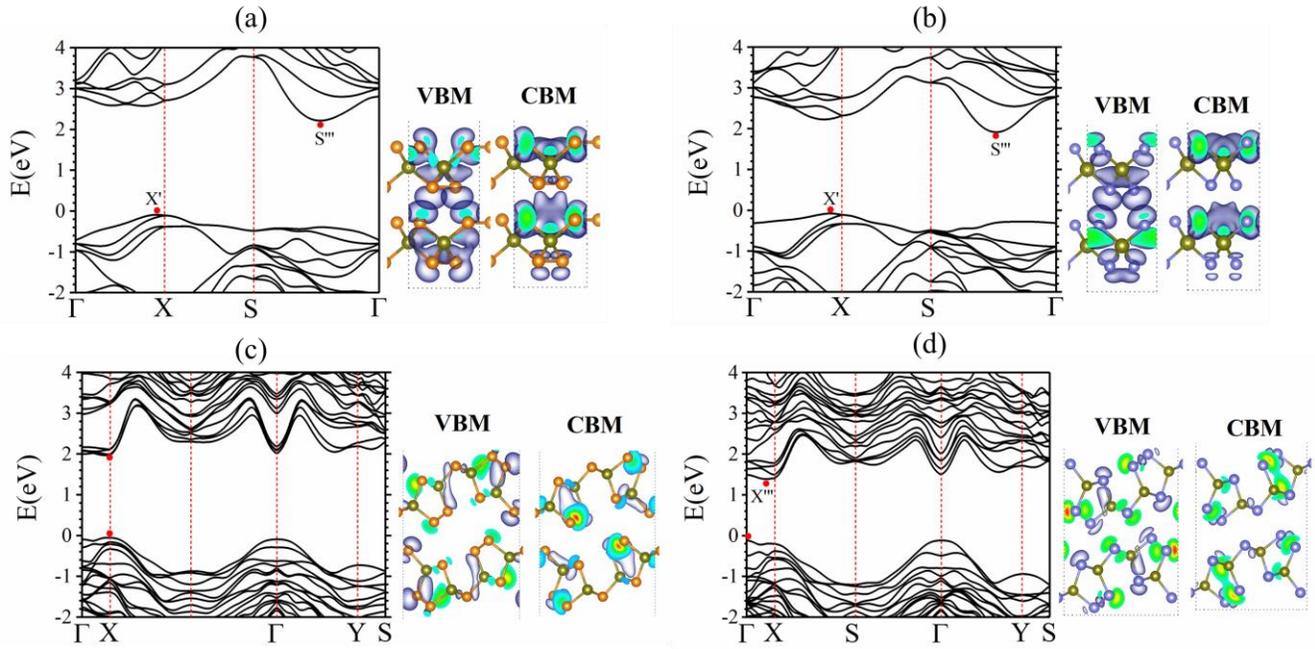

**Figure 6** HSE06 band structures for (a) penta-SiP$_2$, (b) penta-SiAs$_2$, (c) rec-SiP$_2$, and (d) rec-SiAs$_2$ bilayers. Charge density distributions of VBM and CBM are also shown for each bilayer.

To understand the observed trends we carefully analyzed the charge density distributions at VBM and CBM of SiX$_2$ (X = P, As) bilayers. For the three bilayers with strong quantum confinements, the CBM represent a bonding interaction between the CBMs of the monolayers, while the VBM represent anti-bonding interactions between the VBMs of the two monolayers. As a consequence, the CBM is shifted downward and the VBM is shifted upward with respect to those of the monolayer resulting in the decreased band gap. As it can be seen in Fig. 6 (c), due to the smaller size of P atom compared to As atom as well as the corrugated structure of rec-SiP$_2$, no appreciable interaction is observed between the VBMs and CBMs of the two monolayers resulting in small band gap change with respect to that of the monolayer.

In addition to an appropriate band gap, a good photocatalyst is also expected to possess high carrier mobilities so that the photogenerated electrons and holes can participate in chemical reactions before they recombine. Using the effective mass approximation and the deformation potential theory, we next studied the carrier (electron and hole) mobility of SiX$_2$ monolayers along X and Y directions. The calculated elastic modulus ($C_{2D}$), effective mass ($m_i^*$), deformation potential constant ($E_j^1$), and carrier mobility ($\mu_i$) of SiX$_2$ monolayers are given in Table.2. As penta-SiX$_2$ monolayers are structurally isotropic, we only report results for X direction. It has been found that penta-SiP$_2$ exhibits an extremely high electron mobility of 102144 cm$^2$V$^{-1}$s$^{-1}$ , very close to an experimental value for graphene (100000

cm$^2$V$^{-1}$s$^{-1}$), [9] while penta-SiAs$_2$ shows an appreciably smaller electron mobility of 208 cm$^2$V$^{-1}$s$^{-1}$, which is almost the same as that of MoS$_2$ monolayer. The calculated hole mobilities are 796 cm$^2$V$^{-1}$s$^{-1}$ for penta-SiP$_2$ and 503 cm$^2$V$^{-1}$s$^{-1}$ for penta-SiAs$_2$. Due to anisotropy in structure and in turn the electronic band structure, rec-SiX$_2$ monolayers exhibit strongly direction-dependent carrier mobilities. The electron mobility of rec-SiP$_2$ along X direction is found to be 2117 c cm$^2$V$^{-1}$s$^{-1}$, which is about 10 times larger than that along the Y direction (222.71 cm$^2$V$^{-1}$s$^{-1}$). An opposite trend is observed for the hole mobility of rec-SiP$_2$ monolayer: the hole mobility along the X direction is 8 cm$^2$V$^{-1}$s$^{-1}$, which is about 1240 times smaller than that along the Y direction (9636 cm$^2$V$^{-1}$s$^{-1}$). In other words, the electron transport occurs highly favors the X direction, while holes almost exclusively moves along the Y direction. According to the table, rec-SiAs$_2$ monolayer exhibits ultra-high electron mobility of 131881 cm$^2$V$^{-1}$s$^{-1}$ and hole mobility of 7977 cm$^2$V$^{-1}$s$^{-1}$ along the Y direction. These values are 360 and 230 times larger than those along the X direction. The huge difference between the mobilities of electrons and holes in penta-SiP$_2$ and rec-SiAs$_2$ monolayers and highly directional electron and hole mobilities in rec-SiP$_2$ result in an effective separation of photogenerated electrons and holes, which makes them promising light harvesting materials in photovoltaic cells and photocatalysis.

**Table 2** Elastic Modulus (C$_2$D), effective mass (m$_e$*, m$_h$*) of electrons and holes with respect to the free-electron mass (m$_0$), deformation potential constant of CBM and VBM (E$^1_{CBM}$, E$^1_{VBM}$), and mobility (μ$_e$, μ$_h$) of electrons and holes along X and Y directions for SiX$_2$ monolayers. The mobilities are given in cm$^2$V$^{-1}$s$^{-1}$ unit. The effective masses and deformation potential constants are calculated by using HSE06 functional.

| SiX$_2$ | | Direction | C$_2$D (J/m$^2$) | Electron | | | Hole | | |
|---|---|---|---|---|---|---|---|---|---|
| | | | | m$_e$* | E$^1_{CBM}$ (eV) | μ$_e$ | m$_h$* | E$^1_{VBM}$ (eV) | μ$_h$ |
| X = P | penta | X | 67.90 | 0.56 | 0.21 | 102144.98 | 0.46 | 2.92 | 796.81 |
| | rec | X | 24.30 | 1.58 | 0.52 | 2117.45 | 1.58 | 5.94 | 8.22 |
| | | Y | 128.00 | 0.20 | 10.44 | 222.71 | 0.83 | 0.55 | 9636.54 |
| X = As | penta | X | 47.50 | 0.42 | 5.23 | 208.83 | 0.50 | 2.85 | 503.65 |
| | rec | X | 23.20 | 0.99 | 1.88 | 339.75 | 0.83 | 4.86 | 29.53 |
| | | Y | 115.20 | 0.18 | 0.50 | 131881.01 | 0.89 | 0.64 | 7977.12 |

In addition to possessing an appropriate large band gap value and high carrier mobilities, the band edge postions (VBM and CBM) of a photocatalyt must straddle the water redox potentials to identify its likely activity for photocatalystic water-splitting. The CBM energy is supposed to be sufficiently higher than the reduction potential of hydrogen evolution reaction (HER)(H$^+$ /H$_2$ , -4.44 eV with respect to the vacuum level at pH = 0), while VBM energy is required to be lower than the potential of oxygen evolution reaction (OER)($H_2O/O_2$ , -5.67 eV with respect to the vacuum level at pH =0). In order to investigate the criterion-validity, we calculated the absolute energy positions of VBM and CBM of monolayers and bilayers of $SiX_2$ using the HSE06 functional and referenced them with respect to the vacuum level. Fig. 7 compares the VBM and CBM energy levels with redox potentials of water splitting in acidic condition (pH = 0). It can be seen that the band edge positions of all systems the monolayers and bilayers except for rec-$SiAs_2$ monolayer and bilayer straddle the half-reactions potentials, suggesting that they can be useful for photocatalyzed water splitting even without applying bias voltage between the two electrodes. The CBM energy of 1L and 2L rec-$SiAs_2$ are higher than the OER potential of -5.67 by 0.09 and 0.37 eV, respectively, suggesting that they can be used as a photocathode for hydrogen generation. It is well-known that the potentials of HER and OER can be tuned by varying the pH according to the Nernst equations of $E^{red}_{H^+/H_2} = -4.44$ eV + pH × 0.059 eV and $E^{red}_{H^+/H_2} = -5.67$ eV + pH × 0.059 eV, respectively. Therefore, we also examined the photocatalyc activity of the sheets in neutral condition (pH= 7). It is apparent from the Fig. 7 that all $SiX_2$ monolayer and 2L rec-$SiAs_2$ can now straddle water redox potentials at pH = 7. However, CBM energy of 2L penta-$SiP_2$ , 2L rec-$SiP_2$ , and 2L penta-$SiAs_2$ lies almost equal to the potential of HER at pH = 7, suggesting that they can work only as photoanodes which without applied bias.

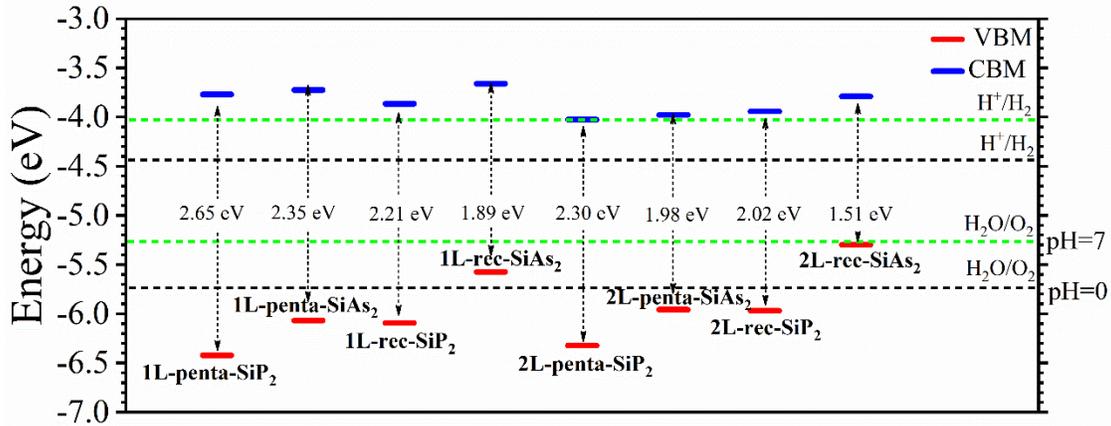

**Figure 7** The band edge positions of the penta-$SiX_2$ and rec-$SiX_2$ monolayers and bilayers. For comparison, the potential of the half-reactions for HER and OER are also shown at two pH values.

To further confirm the feasibility of $SiX_2$ monolayers for the photoconversion, Fig. 8 shows their absorption coefficients computed using HSE06+BSE method. It can be seen that all of the monolayers obviously absorb light in the visible range. The calculated absorption coefficients ($10^5$ cm$^{-1}$) are comparable to those of perovskites, which are known to be highly efficient for solar cells.

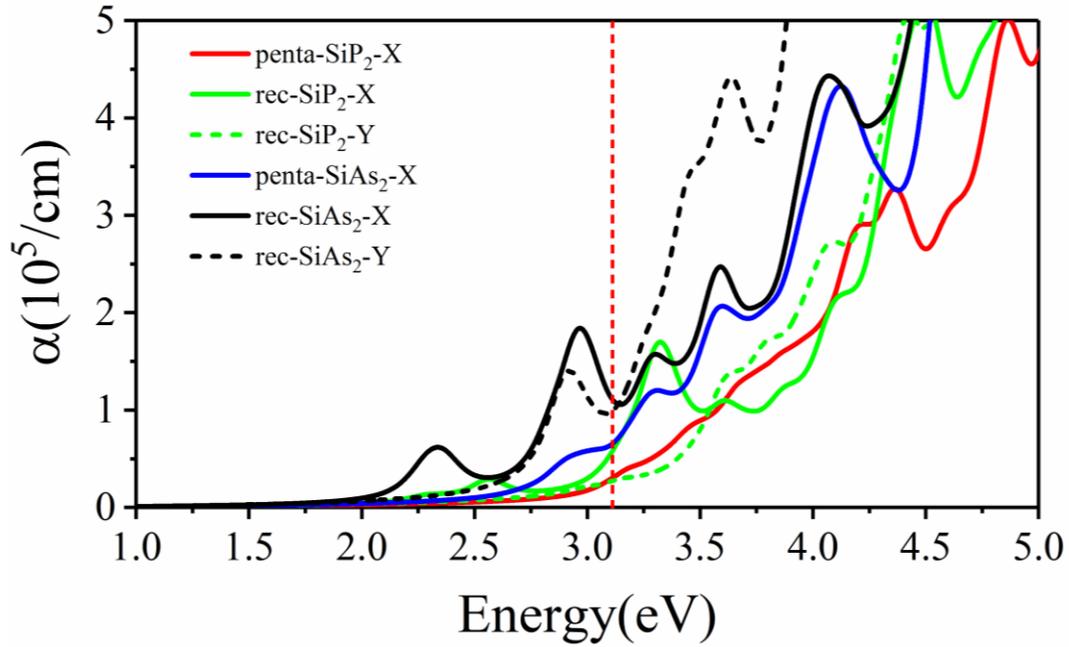

**Figure 8** Frequency dependent absorption coefficients of SiX$_2$ monolayers. The upper edge of the visible light is also shown by vertical dashed line.

## 4. Conclusion

In conclusion, using the extensive first-principles calculation, we studied 2D SiX$_2$ (X = P, As) compounds in pentagonal and orthorhombic phases. HSE06 method results reveal that the monolayer of these 2D materials are all semiconductors with suitable band gaps for photocatalytic water splitting reaction, in the range of 1.89 eV to 2.65 eV. The exfoliation of monolayers from the bulk counterparts is predicted to be experimentally feasible, due to the existence of relatively weak van der Waals interactions between the layers. It is moreover confirmed that the predicted nanosheets are mechanically, dynamically and thermodynamically stable. Notably, these nanosheets are found to exhibit appropriate band edge positions, highly desirable for the water redox potentials. Our calculations confirm the influence of quantum confinement in these 2D materials, resulting in a relatively smaller band gaps of the bilayers in comparison with the corresponding monolayers. The highly anisotropic carrier (electron and holes) mobilities, can yield in an effective separation of photogenerated electrons and holes, which suggest them as outstanding light harvesting materials in photovoltaic cells and photocatalysis. The calculated absorption coefficients (in the order of 10$^5$ cm$^{-1}$) are comparable to those of perovskites, which are known to be highly efficient for solar cells.

**Acknowledgement**

This work is partially supported by Institute for Research in Fundamental Sciences (IPM), Tehran, Iran and the Iran Science Elites Federation. B. M. and X. Z. appreciate the funding by the Deutsche Forschungsgemeinschaft (DFG, German Research Foundation) under Germany's Excellence Strategy within the Cluster of Excellence PhoenixD (EXC 2122, Project ID 390833453).

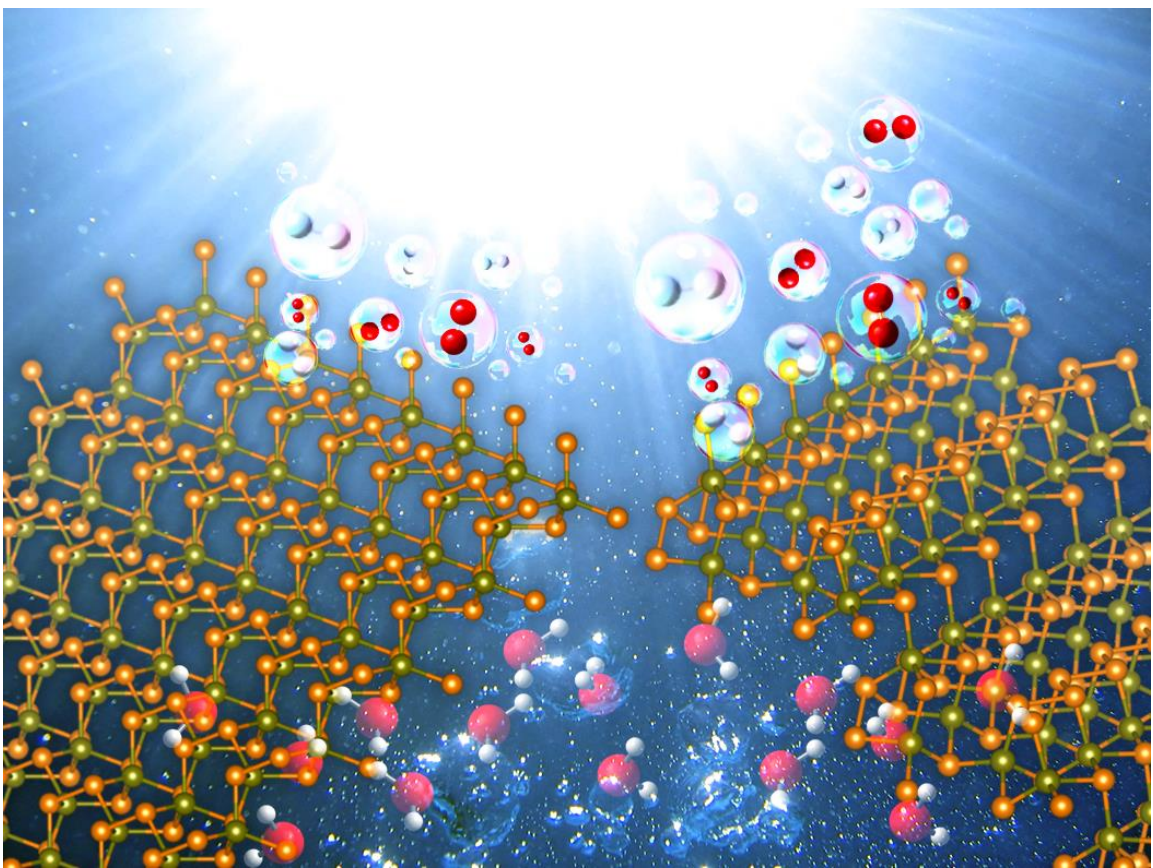

TOC